\documentclass[aps,pra,preprint,amsmath,amssymb,twocolumn,10pt,showpacs]{revtex4}
\usepackage{mathrsfs}
\usepackage{stmaryrd}
\usepackage{verbatim}
\begin{document}
\author{W.\ De Baere}
\email{willy.debaere@UGent.be}
\affiliation{%
Unit for Subatomic and Radiation Physics,
Laboratory for Theoretical Physics,
State University of Ghent,
Proeftuinstraat 86,
B--9000 Ghent, Belgium}


\newcommand{\blt}{$\bullet$}
\newcommand{\com}[2]{\left[#1,#2\right]}
\newcommand{\coms}[2]{\bigl[#1,#2\bigr]}
\newcommand{\hsp}{\hspace{2.5mm}}
\newcommand{\hw}{\wedge}
\newcommand{\Lra}{\Longrightarrow}
\newcommand{\lra}{\longrightarrow}
\newcommand{\me}{\mathscr E}
\newcommand{\mv}[1]{\left< #1 \right>}
\newcommand{\mvs}[1]{\bigl< #1 \bigr>}
\newcommand{\one}{\mbox{{\sf 1}\hspace{-0.20em}{\rm l}}}
\newcommand{\op}[1]{\widehat{#1}}
\newcommand{\scpr}[2]{\left< #1| #2\right>}
\newcommand{\mel}[3]{\left< #1\left| #2\right| #3\right>}
\newcommand{\ket}[1]{\left| #1\right>}
\newcommand{\bra}[1]{\left< #1\right|}

\title{
On the Consequences of Retaining the General Validity of
Locality in Physical Theory}

\begin{abstract}
The empirical validity of the locality (LOC) principle of relativity is
used to argue in favour of a local hidden variable theory (HVT) for
individual quantum processes. It is shown that such a HVT may reproduce
the statistical predictions of quantum mechanics (QM), provided the
reproducibility of initial hidden variable states is limited. This means
that in a HVT limits should be set to the validity of the notion of
counterfactual definiteness (CFD). This is supported by the empirical
evidence that past, present, and future are basically distinct. Our
argumentation is contrasted with a recent one by Stapp resulting in the
opposite conclusion, i.e.\ nonlocality or the existence of
faster--than--light influences. We argue that Stapp's argumentation
still depends in an implicit, but
crucial, way on both the notions of hidden variables and of CFD.
In addition, some implications of our results for the debate
between Bohr and Einstein, Podolsky and Rosen are discussed.
\end{abstract}
\maketitle

\vspace*{0.5cm}

\section{INTRODUCTION}
\label{introduction}
Bell's inequality (BI)\cite{1} illustrates the
fundamental conflict between the quantum formalism and any formalism,
aimed at reproducing QM, in which the validity of {\em all} classical
principles is assumed in a quantum context (the so--called ``Hidden
variable (HV) program" according to Kochen and Specker \cite{2}). In
spite of the existence of very reasonable arguments -- both theoretical
and empirical -- which retain the validity of the relativistic locality
principle,\cite{3,4,5,6,7} and
which remove any quantum inconsistency or ``paradox", some people continue
to propose arguments in favour of nonlocality (for a survey of the recent
literature, see e.g.\ Ref.\ 8).

One sequence of such nonlocality argumentations has been set up by H.P.\
Stapp, starting in 1971\cite{9} and so far ending with his
recent paper in Am.\ J.\ Phys.\cite{10} While his earlier
argumentations were based explicitly on the universal validity of CFD and
on the assumption of the existence of hidden variables (HVs), his later
ones are claimed to be no longer dependent on these assumptions.
Stapp concludes that the premises of QM entail
``\ldots {\em some sort of failure} of the notion that no influence
{\em of any kind} can act over a spacelike interval.''

In his last nonlocality proof, Stapp\cite{11} is concerned to
compare ``\ldots the possible consequences of making different
choices\ldots" and admits that ``\ldots the argument involves a certain
weak form of counterfactual reasoning." It was precisely the use of CFD in
this work that was criticized by Mermin,\cite{12} Unruh,\cite{13}
and Shimony and Stein.\cite{14} As on
past occasions (see Refs.\ 15,16,17) Stapp could
not be convinced, as these criticisms were followed by several
replies.\cite{18,19} In the Abstract of
Ref.\ 10 it is stated
again: ``The premises include neither the existence of hidden variables
nor counterfactual definiteness, nor any premise that effectively entails
the general existence of unperformed local measurements".
In the same paper Stapp admits once more that
``The argument given above rests heavily upon the use of counterfactuals:
the key statement $SR$ involves, in a situation in which $R2$ is performed
and gives outcome `+,' the idea `if, instead, $R1$ had been
performed...'". In Ref.\ 11 Stapp complains about the
criticism that his nonlocality argumentations still depend on
counterfactual reasoning: ``This fact is sometimes taken as a sufficient
reason to discard wholesale all EPR--Bell--type arguments. That tactic is
not rational." Yet, in former work Stapp\cite{20} stated that
``No satisfactory derivation of nonlocality, or the existence of
faster--than--light influences, can be based upon such a CFD assumption: a
failure of this assumption is (at least in my opinion)
far more likely than the existence of a faster--than--light influence.''
From Stapp's persistence to prove the nonlocality property of QM, it is
apparent that he has changed his opinion considerably, and tries to remove
the assumption of CFD from his nonlocality proofs. Because also his recent
proof needs CFD in some (stronger or weaker)
form, the previous criticism
applies also to that new proof.

In the present work our criticism of that kind of
proofs is also directed against the unjustified use of CFD. The
difference, however,
with the work by Mermin, Unruh, and Shimony and Stein, is that
we are able to give a plausible
physical explanation for the possible
invalidity of CFD in the context of all nonlocality proofs. Our approach
is based 1) on the argument that for entertaining reasonable and rational
argumentations in which {\em single} quantum events are involved, one
should first assume {\em explicitly} the existence of an appropriate
quantitative
formal scheme describing such processes, and 2) on the further
obvious requirement that the theory may not violate basic empirical data,
such as Einstein locality as specified in Sec.\ 2.

In Ref.\ 10 Stapp then continues by arguing that his use
of CFD is justified (or at least is different from the use of CFD by e.g.\
Einstein, Podolsky and Rosen (EPR), and by Bell in a HV context) on the
ground of his assumptions ``1.\ Free Choice", ``2.\ No Backward in Time
Influence", ``3.\ Validity of Predictions of quantum theory (QT): Certain
predictions of quantum theory in a Hardy--type experiment are valid."
However, we show in Sec.\ 6 that the context in which his
assumption ``1.\ Free Choice" is used, is equivalent with the assumption
of HVs. Furthermore, the passages in Stapp's proof
which deal with alternative choices, definitely concern {\em individual}
events, and are all of the counterfactual type.
In order to show the unjustified use of CFD in these nonlocality--of--QM
proofs, we will specify in Sec.\ 5
in detail the conditions which
justify counterfactual considerations.
This will allow us to show in Sec.\ 6
the equivalence
between the assumptions ``Free Choice" and HVs.

Although it is widely believed that it is the principle LOC that should
break down (i.e.\ Bell's theorem), we continue our past
approach\cite{3,4,15,16,17} and argue that the origin
of the conflict between QM and classical--mechanics (CM)--like
schemes is the assumed general
validity of the idea of CFD, applied to the
domain of {\em individual} quantum processes. At first sight this
conclusion coincides with the critical remarks by Mermin and Unruh
(who warn to be careful with the use of
CFD), and with those by Shimony and Stein (in terms of ``possible
worlds"). There are, however, three main differences with our procedure to
get our conclusion:
\begin{itemize}
\item[1)] we point to the fact that
any nonlocality--of--QM argumentation (including Stapp's recent one)
intermingles valid QM predictions
with predictions for
{\em alternative individual incompatible}
measurement outcomes;
\item[2)] we remark that
in so doing one leaves the domain of
validity and of applicability of QM (which is about {\em ensembles} of
similarly prepared systems), and consistency demands the introduction of a
quantitative formalism for these {\em individual} quantum processes, i.e.\
the existence of something like a HV formalism is all the time implicitly
presupposed. The alternative, followed in most nonlocality argumentations,
is to deny the need of HVs and to
recourse in unreliable semantics;
\item[3)]
we refer to former work by Houtappel, Van Dam, and Wigner,\cite{21}
who stress the equal importance of initial conditions
within a particular theory. Joining this view, we are able to give
a plausible physical significance of
a possible invalidity of CFD at the individual quantum level of
description.
First, it means that the breakdown of CFD concerns the HVT, and not QM
-- in QM CFD {\em is} valid, see Sec.\ 5. Second,
the HV states which represent the initial conditions within
the HVT, do no longer have the property of being reproducible in future
preparations. In former work we called this the ``nonreproducibility
hypothesis" (NRH);
\item[4)]
our conclusion NRH points, possibly, to the origin of the existence of
incompatible observables in QM,
and, maybe, to the existence of the
``arrow of time" as a basic property of nature's evolution.
\end{itemize}

\section{THE PRINCIPLE OF LOCALITY}
\label{deflocality}
The requirement  that
 ``\ldots no influence
{\em of any kind} can act over a spacelike interval'' may be considered as
a general definition of locality. Its quantitative expression depends
on the theory which describes these influences or physical processes. If
this influence is understood as pertaining to e.g.\ quantized fields, one
gets the condition that the field operators associated with
spacelike separated regions should commute (see Sec.\ 3.3).
Ultimately, these fields are used to make statements about measurement
outcomes.
Because the assumed causal relationship between the sets of
past and future outcomes, is also
the result of the propagation of these
physical influences, the above locality principle applies equally to
these observable quantities.
So, because these observations may be
considered as ``events'' in space--time,
the locality principle as defined above, includes the locality principle
of relativity. According to Einstein,\cite{22} it
means that ``\ldots the real, factual situation of $S_1$ does not depend
on what is done with $S_2$ which is spatially separated from the former''.
This locality principle has been taken over by
Bell\cite{1}
(``\ldots the result of a measurement on one system be
unaffected by operations on a distant system with which it has interacted
in the past\ldots''), and is the one adopted in most other papers.
Formulated in this way, the locality principle concerns observable
quantities, and is called also ``Einstein locality''. It is this
definition of LOC which will be adopted in the present paper.
In QM these observable quantities are
represented by Hermitean operators, and in a tentative HVT the
representation is e.g.\ by ``result functions'' $A,A',B,B'$, i.e.\ the
ones figuring in Bell--type inequalities and in nonlocality--of--QM
argumentations.

\section{ARGUMENTS IN FAVOUR OF LOCALITY}
\label{loc}
Successful and acceptable physical theories
are built on principles which
correspond with well established basic empirical data.
In the case of QM, wave--particle
duality and Einstein locality, as defined in Sec.\ 2,
belong to these data.
The first
one corresponds with the superposition principle, and the second one
guarantees the independence of the statistical QM predictions from actions
or choices in far--away regions, outside the observer's
light cone.\cite{23,24}
As a result, these principles constitute the basis of
all existing (quantum) theories, and no violation of these principles have
ever been observed.\footnote{The latest experiment being that of
Stenner et al.,\cite{25} see also the reports by
Steinberg.\cite{26,27}}
Therefore, in our view it
seems rational to accept (instead of to criticize) the validity of LOC
within QM and in any other physical theory. It is our purpose to convince
the reader of the overwhelming evidence in support of LOC. In
addition to the empirical evidence, we summarize below some
further theoretical arguments in favour of the validity of LOC.

\subsection{Nonlocality Cannot Have General Validity}
Consider one single individual correlated quantum process. Observations
confirm that the evolution between the preparation and each of the
correlated measurements proceeds in a local way, in agreement with
relativity. On the one hand, according to the idea of quantum nonlocality
-- or at least according to some of its interpretations --, the
simultaneous correlated measurement processes should influence each other
in a nonlocal way. Until now no one has been able to give a reasonable
explanation for this kind of dichotomy in the evolution of physical
processes. However, this problem can be removed immediately by
accepting the general validity of locality, and by considering the
nonlocality argumentations as evidence for
other nonclassical
properties of underlying HV states, such as our NRH conclusion.
This removes all contradictions, and
everything is now again
supported by empirical data.

\subsection{QM is Formally Local}
In general, QM makes deterministic predictions only
for the expectation values of observables.
It follows that in QM it is only via such
deterministic statistical predictions that a nonlocal QM property may,
eventually, be observed.
In this respect,
it has been shown by Ghirardi, Rimini and Weber\cite{23,24} and by
others, that these predictions are independent of far--away actions.
Hence, statements -- or conclusions -- about the nonlocal influencing of
{\em single} measurement outcomes by other, far--away, operations cannot
be the result of argumentations held within QM. Such
argumentations require rather
the introduction of another formalism, in terms of
theoretical elements or notions not contained in QM.

\subsection{Locality is a Basic Principle in all Elaborations of QM}
\label{elabqm}
Locality is never empirically violated and, as a result, it is built in
{\em explicitly} in all successful physical theories,
such as in any formulation of relativistic quantum
field theory (QFT), string theory, etc.
In these theories, field operators
$\op{A}(x)$ and $\op{B}(y)$
in space--time points $x,y$
satisfy the commutation relation\cite{28}
\begin{equation}
[\op{A}(x),\op{B}(y)] = i \Delta(x - y) = 0 \hspace*{.5cm}{\rm
for}\hspace*{.5cm} (x - y)^2 < 0\;,
\label{microcausal}
\end{equation}
known as ``microscopic causality" or ``local commutativity.
In (\ref{microcausal}), $\Delta(x-y)$ is
a singular function vanishing outside the light cone. This basic relation
represents the universal validity of causality and locality. It means that
for spacelike separated intervals $x-y$, the observables $\op{A},\op{B}$
commute and, hence, can be measured together in such a way that the
processes resulting in the measured outcomes do not disturb each other,
i.e.\ are independent, in agreement with the principle LOC.
Similarly, in Weinberg's
classic book\cite{29}
the commutativity of spacelike separated observables
$\psi_l(x),\psi_{l'}(y)$ is stated as:
\begin{equation}
\coms{\psi_l(x)}{\psi_{l'}(y)}_{\mp}=
\coms{\psi_l(x)}{\psi^{\dagger}_{l'}(y)}_{\mp}=0
\label{weinbergp198}
\end{equation}
with the justification ``\ldots The condition (\ref{weinbergp198})
is often described as a {\em causality}
condition, because if $x-y$ is space--like then no signal can reach $y$
from $x$, so that a measurement of $\psi_l(x)$ at point $x$ should not be
able to interfere with a measurement of $\psi_{l'}$ or
$\psi^{\dagger}_{l'}$ at point $y$.''
For an axiomatic approach, see e.g.\ Ref.\ 30.

 From the success of these theories,
it may be concluded that there is no reason why the conflict between QM
and the BI should be interpreted as a breakdown of LOC in QM. Recently,
the validity of locality and causality in QFT has been defended by
Tommasini.\cite{31}

\subsection{Irrelevance of Nonlocality in the Conflict between QM and
Classical--like Theories}
\label{conflqmcm}
The idea of quantum nonlocality results from a
program which tries to reconstruct QM (which makes deterministic
predictions for {\em statistics}) from a more detailed theory of the
classical type for the {\em individual} case. According to Kochen and
Specker\cite{2} this program investigates ``\ldots the possibility of
embedding quantum theory into a classical theory \ldots". However,
this HV program is ill defined because it violates
basic QM rules from the outset.\cite{32,33} Formally this is
evident from Landau's identity:\cite{34}
\begin{equation}
\bigl(\op{A}\op{B} + \op{A}'\op{B} + \op{A}\op{B}' -
\op{A}'\op{B}'\bigr)^2 = 4\one + [\op{A},\op{A}'][\op{B},\op{B}']
\label{landau}
\end{equation}
where the observables $\op{A}$, $\op{A}'$, $\op{B}$, $\op{B}'$ represent
correlated measurements along 4 possible, but incompatible,
experimental settings:
$a,a'$ in one region $L$ and $b,b'$ in another
spacelike separated region $R$. From Eq.~(\ref{landau}) it follows that
the expectation value of the left hand side satisfies Cirel'son's
inequality:\cite{35}
\begin{equation}
\bigl| \mvs{\op{A}\op{B}} + \mvs{\op{A}'\op{B}} + \mvs{\op{A}\op{B}'} -
\mvs{\op{A}'\op{B}'}\bigr|\; \leq 2\sqrt{2}\;.
\label{cirelson}
\end{equation}
It is worthwhile to remark here that Eq.\ (\ref{landau}), and hence also
Eq.\ (\ref{cirelson}), is a consequence of property (\ref{microcausal}),
i.e.\ of the assumed commutativity of spacelike separated QM observables,
 being an immediate formal expression of Einstein locality.
The fact that empirical data do not violate Eq.\
(\ref{cirelson}), may be considered as convincing evidence that
the principle of Einstein locality is firmly incorporated in QM,
at least
if one restricts to its domain of validity and of application, namely the
ensemble of similarly prepared individual cases. Leaving this domain
easily leads to wrong conclusions, such as the empirically not
supported quantum nonlocality.
In this respect one may apply
van Kampen's ``theorem IV",\cite{36}
according to which ``\ldots whoever endows $\psi$ with more
meaning than is needed for computing observable phenomena is responsible
for the consequences" -- here, for explaining how the
alleged quantum nonlocality is compatible with the empirical validity of
Einstein locality.

Now, in order to get a BI from Eq.\ (\ref{cirelson}),
a further assumption should be added, namely the validity of CFD
for {\em single} quantum processes.
Restricting to quantitative considerations,
this brings us back to the HV program.
The assumed validity of CFD (either
explicitly or implicitly) in any HV program,
amounts in fact to a return to the
classical formalism in which all observables are
compatible and all commutators vanish. Otherwise stated, such a
program violates QM already from the start, not because of
a violation of locality, but
because of the assumed unrestricted validity of CFD for {\em single}
quantum events. QM {\em and} locality are valid because the data satisfy
Cirel'son's relation (\ref{cirelson}).

In such classical--like HV programs,
inequality (\ref{cirelson}) reduces to the BI:
\begin{equation}
\bigl| \mvs{\op{A}\op{B}} + \mvs{\op{A}'\op{B}} + \mvs{\op{A}\op{B}'} -
\mvs{\op{A}'\op{B}'}\bigr|\; \leq 2,
\end{equation}
which is violated by QM predictions.
It is seen that it is the nonvanishing of local, single
particle's, commutators $[\op{A},\op{A}']$ and $[\op{B},\op{B}']$,
which is responsible for the violation of BI by QM.
This leads to the conclusion that it is the noncommutativity of
quantum observables pertaining to a single
system (which may be part of a larger one of correlated systems)
that plays
a crucial role in the violation of the BI by QM.
Hence,
nonlocality cannot be held responsible for
that violation.

It is our opinion that this extended list of arguments in favour
of locality, is convincing enough to allow the conclusion that the
conflict between BI and QM may only be explained by rejecting the
other -- mostly implicit --
basic assumption, namely the validity of CFD at the subquantum level.

\section{HIDDEN VARIABLES AND SINGLE QUANTUM PROCESSES}
\label{phystheory}
We argue in this section
that all nonlocality argumentations contain statements
about individual quantum processes, and
that in general hidden variables are needed for their quantitative
description.

\subsection{The Structure of Physical Theory}
\label{pragmaticth}
We assume that physical processes proceed in a lawful
way, and that the relevant basic laws have a mathematical expression. Our
physical theories, then, are representations of these basic laws, and we
shall consider them as {\em deterministic} formal schemes, which allow an
observer to connect in a mathematical way causal relationships between
consecutive observations. The knowledge about the first set of
observations may be considered as initial conditions within the physical
theory. The extension of this formal scheme with statements about the
significance of the theory's concepts with respect to an eventually
underlying reality, belongs to the interpretation of the theory. In this
work, however, we shall not go beyond the formalism itself, i.e.\ we take
on a pragmatic attitude. The main point is that the theory is considered
as a procedure to make predictions for future observations, given enough
knowledge of previous ones. In this approach notions such as ``reality",
``realism", ``assigned values", etc.\ gain their significance only by
their mathematical representation. E.g.\ ``reality" is represented by the
theory's state, ``assigned value" by the prediction of a measurement
outcome, etc.

In any formal scheme a physical situation is
represented by a state function, and the
space--time evolution is determined by dynamical
equations. Because the theory concerns observations by
macroscopic observers, a procedure should be present for
relating measurement outcomes with the physical state. Another
important notion (yet frequently overlooked in foundational issues)
is that of the domain of
applicability and of validity of the theory. This is the domain
for which the theory makes verifiable, hence deterministic,
predictions for measurement outcomes in a successful way. As a
consequence of our pragmatic view, we
will consider any statement in an
argumentation as a valid and a justified one, only if that statement can
be verified in an {\em actual} experiment. This is an obvious
(yet frequently overlooked) requirement,
because otherwise there is no means to decide whether
a statement is false or true. This guiding rule will be used
in Sec.\ 5
to set up criteria for justifying counterfactual statements.

Let us now look back at CM, QM and HVT from the above
pragmatic point of view.
CM's domain is the description of single processes
(giving rise to macroscopic observations) belonging to the
macroscopic domain. QM's domain, on the contrary, is the
description of the ensemble of single microscopic processes
(again giving rise to macroscopic outcomes). Only for such an
ensemble are the predictions determined in a deterministic, hence
verifiable, way. It follows that if considerations are
held for processes which fall outside a theory's
(e.g.\ CM's or QM's) domain, then this implies the implicit
assumption that the reasoning presupposes
another theory which is appropriate for the
situation considered. In particular, this is the case with
all nonlocality--of--QM argumentations,
in which outcomes of single {\em actual}
quantummechanical measurement processes are always compared with
other single, but incompatible, {\em hypothetical} outcomes, i.e.\
outcomes which ``would have occurred if the alternative,
incompatible, choice had been made".

We conclude that
in all nonlocality--of--QM argumentations
the availability of a HVT for single quantum processes
is presupposed.
It is only under the further assumption,
that the HVT's initial conditions may be reproduced, that
it is justified to compare predictions for single outcomes.
If it turns out, finally,
that inconsistencies arise,
then one reasonable conclusion is
that the HVT's initial conditions cannot
be actualized, i.e.\ reproduced, again.

Our point is that in many of these argumentations
the need of HVs or a HVT is denied.
Instead, the hypothesis of HVs or a HVT is replaced by a semantical
description in terms of words only. In this way the assumption of a HVT is
hidden and made implicity. To remedy for this
inconsistency, the idea of a HVT must be introduced explicitly.

Also in his recent work Stapp rejects a priori an analysis in terms of
HVs, on the incorrect grounds 1) that the HV assumption is ``\ldots
logically equivalent to the assumption that values can be pre--assigned
conjunctively and locally to all of the outcomes of all of the alternative
possible measurements.", and 2) that it ``\ldots conflicts with \ldots the
orthodox quantum philosophical attitude that one should not make any
assumption that effectively postulates the existence of a well defined
outcome of a localized measurement process that is not performed."
Argument 1) is incorrect because from the end of the previous paragraph
and from Sec.\ 5 it follows that it can be invalidated
easily by setting limits to the reproducibility of initial states in the
HVT -- which amounts to setting limits to CFD. Argument 2) is incorrect
because QM has in general {\em nothing at all} to say about ``the
existence of a well defined outcome of a localized measurement process
that is not performed", i.e.\ in QM such an assumption can neither be made
nor be denied a priori on ``orthodox quantum pilosophical" grounds. Only
when a new formalism, such as a HVT, is presumed does it have sense to
ponder about arguments 1) and 2).

Applying, then, the format of CM and QM to the hypothetical HVT, a
general state
$\lambda(x,t)$ should be introduced as a representation of the individual
physical situation (or, in EPR's terminology, a time--dependent ``element
of reality"),
in the same way as the QM state $\ket{\Psi}$
is a representation of the ``reality corresponding with the ensemble".
This individual situation
may well be the result of an observer having
made a particular choice. But {\em different} choices cannot be considered
together, because only one of them may correspond to a measurement process.
Hence, different choices for {\em individual} processes should be
represented by different HV states.

As in CM and QM, dynamical laws determine in a deterministic way the HV
state at any other instant. Because the HVT is aimed to describe
individual situations, the HV state $\lambda(x,t)$ may be considered as a more
faithful representation of nature -- or reality -- than the QM state
$\psi(x,t)$.
Finally, a link with observation by a conscious observer is
established by introducing result functions $R(\lambda_A(x,t))$, which give
the outcome $r=R(\lambda_A(x,t))$ when the observable $A$ is measured.

\subsection{The Role of Initial Conditions}
Houtappel, Van Dam, and Wigner\cite{21} stressed
the equal importance of the initial conditions and of the laws of
physics ``\ldots because the laws of nature do not lead to
observable consequences unless the initial conditions are given
\ldots". These initial conditions concern the theory's state
representing, in terms of EPR's terminology, some ``element of
physical reality".
In CM, the initial state is determined by the
prepared values of a number of observables (which all are
compatible, the prepared value of one observable not being
influenced by the preparation of any other one). In QM the initial
state is determined by the  some repeated
preparation procedure, and in a HVT the prepared initial state is
represented by $\lambda(x^{\mu})$, which is up to now completely
unspecified and, hence, very general. Whereas in CM and in QM
states may be reproduced, the reproducibility of
$\lambda(x^{\mu})$ (which would justify the use of CFD in HVT,
see Sec.\ 5)
should have the character of an extra assumption.

\subsection{Determinism or Indeterminism?}
\label{determinism}
Determinism is a formal property of any theory.
It is this property which allows the theory to make verifiable
predictions. In the case of QM,
the deterministic predictions concern the
statistics of outcomes within an ensemble of measurement events.
Because the statistics obey deterministic laws, it seems plausible to
assume that also the individual quantum processes obey deterministic laws.
We consider this a reasonable approach (among other reasonable ones).

One
of the origins of the dissatisfaction with QM is precisely the discrepancy
between the fact that QM in general makes only statistical predictions in
a deterministic way, and the idea that the observed individual outcomes
should allegedly have no determinite cause in the past (such as in
Copenhagen--like interpretations of QM\cite{37}). In the words of
Stapp:\cite{38} ``Some writers claim to be comfortable with the
idea that there is in nature, at its most basic level, an irreducible
element of chance. I, however, find unthinkable the idea that between two
possibilities there can be a choice having no basis whatsoever. Chance is
an idea useful for dealing with a world partly known to us. But it has no
rational place among the ultimate constituents of nature." Such words are
reminiscent of Einstein's view that ``God does not play dice".
However, on account of Stapp's recent attempts to prove
the nonlocality property of QM,
this former view of him is clearly opposite to his present one.

An apparently overlooked -- or forgotten --
old interferometer thought experiment by Renninger,\cite{39}
may be invoked as an indication for
the existence for determinism even at the {\em ontological} level. Indeed,
in Renninger's thought experiment, it is possible to influence in a causal
way the realities corresponding with a single quantum system moving in
both arms of an interferometer, in such a way that the final outcome is
predictable with certainty, no matter how many times --
in principle an infinite number of times --
one has influenced these realities moving simultaneously in both arms.

The deterministic laws for the individual case, then, constitute the HVT
for individual quantum processes. At present, however, the observer's
limited technical capabilities prevent control over the supplementary
observables. Such a HVT should then bear the general properties of a
physical theory, discussed in Sec.\ 4.1. This means that
also in a HVT, locality should still be a valid principle, and that for
given initial conditions (of which one is not {\em a priori} sure that
they are reproducible, as is yet empirically the case in CM and QM) of the
HVs, the HVT should predict in a deterministic way the unique outcome of
the future event that will actually happen.

Some further remarks with respect to determinism may be made here.
Jammer\cite{40}
notes that ``It has been claimed that even the
most ``progressive" theoretician believes at the bottom of his heart in a
strictly deterministic, objective world even if his teachings
categorically deny such a view \ldots It explains, however, why some
physicists rejected the prevailing probabilistic interpretation of quantum
mechanics and tried to demonstrate that the existing theory in spite of
its spectacular success is only a provisional approximation to a deeper
scientific truth." With respect to the realizability of EPR's program
(suggesting a more detailed, deterministic,
description of the individual quantum process)
Dirac's opinion\cite{41} is that ``\ldots we think it might turn out
that ultimately Einstein will prove to be right, because the present form
of quantum mechanics should not be considered as the final form. There are
great difficulties \ldots in connection with the present quantum
mechanics. It is the best that one can do till now. But, one should not
suppose that it will survive indefinitely into the future. And we think it
is quite likely that at some future time we may get an improved quantum
mechanics in which there will be a return to determinism and which will,
therefore, justify the Einstein point of view". More recently, in an
interview on Dutch TV, S.\ Weinberg expressed a similar opinion and
claimed that maybe, finally, natural processes proceed in an entirely
deterministic way. It may be noted also that recently
't Hooft\cite{42} started developing deterministic models for individual
quantum processes.

\section{CRITERIA FOR A JUSTIFIED USE OF CFD}
\label{criteriacfd}
It is argued in the present work, that the
reason for the appearance of quantum contradictions is the {\em
unjustified} use of CFD in the domain of {\em individual quantum}
phenomena.
Usually, the general validity of CFD is accepted either
as a self--evident assumption, or by
invoking the authority of Bohr: ``Bohr repeatedly emphasized the freedom of
experimenters to examine either aspect or another of an individual
quantum system" (Stapp in Ref.\ 11, p.\ 301), or:
``\ldots Bohr did not want to take the difficult road of trying to ban
all use of counterfactual concepts in physics \ldots
he recognized that counterfactual concepts do play an important role
in the pragmatic approach to physics that he was pursuing."
(Stapp in Ref.\ 10).

Also others refer to Bohr, e.g.\ Peres\cite{43}
states that ``Bohr did not contest the validity of
counterfactual reasoning. He
wrote: `{\em our freedom of handling the measuring instruments is
characteristic of the very idea of experiment \ldots we have a completely
free choice whether we want to determine the one or the other of these
quantities\ldots}' \ldots Just as EPR, Bohr found it perfectly legitimate
to consider counterfactual alternatives. He had no doubt that
the observer
had free will and could arbitrarily choose his experiments"
(emphasis by A.\ Peres).
We remark here that there is no incompatibility between
a possible determinism on the HV level of description, and
our {\em experience} of free will.
The reason is that the observer's limitations prevent him to know, or
to become aware, of his precise HV state $\lambda(x,t)$.

Both Stapp and Peres argue that in everyday life and in physics, CFD is an
essential part of reasoning. Indeed, it is on account of the validity of
CFD in CM, that scheduling a journey is an allowed reasoning, and in
Ref.\ 11 Stapp considers the case where and electron
```would have landed' if the experimenter had used in the experiment a
second apparatus, `instead of' the first one". The point is that in both
of these CM cases, CFD is used in a justified way, because the initial
conditions may be reproduced, allowing the verification of the
counterfactual statement. According to Stapp this then ``\ldots
illustrates the fact that theoretical assumptions often allow one to say
with certainty, on the basis of the outcome of a certain experiment, what
`would have happened' if an alternative possible apparatus had been used."

It is evident that similar examples may be given also in the domain of
applicability and of validity of QM.
In QM the state $\ket{\Psi}$ is
reproducible and, when the expectation value $\mvs{\op{A}}$ is actually
measured, then it has sense to ask the counterfactual question ``What
would have been the expectation value $\mvs{\op{B}}$ if another observable
$B$ were measured instead''. This counterfactual question has sense
because it can be investigated in a real, actual, experiment
described by the same
quantum state $\ket{\Psi}$. By looking why CFD is valid in the
specific examples of CM and of QM, we get the following three criteria for
a {\em justified} use of CFD in each particular situation.

\subsection{Criterion of Availability of a Theory (CT)}
\label{ct}
The above examples of CM and QM
suggest that an unproblematic use
of CFD is connected with the availability of a working theory, and
with theoretical possibilities which are allowed not only
by that theory, but also by nature itself.
If it turns out that some theoretical possibility
is empirically not allowed, then
such a possibility cannot be considered counterfactually.
For example, in Newton's theory
negative mass is a theoretical possibility which is not allowed by
nature. According to this theoretical possibility,
particles in a graviational field should not fall down but move
upwards, contrary to any observation.
In general, superselection rules are added to the theory ``by hand",
in order to remove such cases or initial conditions.

Now, as a rule
quantum paradoxes generally arise when QM
predictions, dealing with ensembles of quantum events
(for which a detailed theory, QM, is available), are tried to be
reconstructed by means of considerations about the {\em individual}
events of this ensemble (for which as yet {\em no} detailed theory
is available).
However, consistency then demands that such considerations
be held within some theoretical framework, so that
all assumptions may be stated in an {\em explicit} and quantitative
way. We shall call this requirement the criterion CT, i.e.\
for each particular case, a
quantitative formal scheme must always be presupposed.

\subsection{Criterion of Verifiability (CV)}
Physical theories should make verifiable predictions, and this implies
that they are {\em formally deterministic}.
This property
determines the theory's domain of applicability, and as long as the
predictions are correct, the domain of applicability coincides
with the domain of validity.

For instance, CM makes {\em deterministic} predictions for {\em
individual} events, and its domain of applicability is that of
the individual case.
However, CM does not apply to individual {\em quantum} processes,
so that its domain of
validity is rather restricted. Therefore, CM is not complete.

QM, on the other hand, in general makes {\em deterministic}
predictions only for the {\em statistics} of measurement outcomes
within an {\em ensemble} of measurements following a specific
preparation.
It follows that the domain of applicability and
of validity of QM is that of ensembles. Of course,
these deterministic QM
predictions for ensembles of quantum systems, may be transformed
into probabilistic statements about individual
measurement outcomes. However, these probabilistic statements about
individual cases do not satisfy CV. Therefore, the description of the
individual quantum process
does not belong to the domain of applicability of
QM, but rather belongs to its interpretation. On this ground
neither QM may be termed a complete theory, which was also the
conclusion of EPR's argumentation.

Finally, applied to possible HVTs
for individual quantum processes, CV requires them
to be also of the deterministic type. In such HVTs,
individual quantum preparations of both the object system and the
measurement system for a specific observable, should allow then
to make predictions with certainty for that observable.

\subsection{Criterion of Actualizability (CA)}
Given a theoretical scheme,
physical relevance of counterfactual reasoning
requires furtheron that the conditions
which underly physical argumentations not only are {\em
theoretically} possible, but also that they may
correspond with {\em actualizable} physical situations.

In particular, counterfactual argumentations which concern
non--actual processes and events, have physical relevance only
insofar as one is guaranteed of the possible actualizability, {\em
at least once}, of the non--actual process. It should be clear
that such a guarantee cannot come from the allowance by the theory
of a particular situation as a theoretical possibility, but may
come only from actual experience. Such an actualizability or
reproducibility condition would justify not only the use of CFD,
but also the (further) assumption that counterfactually
considered processes
obey the same rules as actual processes.

In summary, counterfactual reasoning in some domain is justified provided
it is assumed that a formally deterministic theory for the relevant
processes exists, and that the physical situations it describes are
actually reproducible, in the sense that they are representable by
identical states within the theory. If this assumption does not correspond
with reality, then a number of situations will be encountered for which
paradoxical results may be derived.

So, it may be envisaged that restricting the justified use of CFD for
single quantum processes, is one of the possibilities to resolve
inconsistencies ensuing from attempts to reconstruct QM in terms of a
classical framework. The physical ground for a possible invalidity of
CFD at the individual quantum level is that CA is not fulfilled, or that a
reproducibility hypothesis (RH) concerning a former actual situation is
not valid.

A picture in terms of nonreproducible situations at the subquantum level
provides, then, an alternative physical explanation for the apparent
``nonlocality" in various correlated situations (of the EPR--Bell--type),
and for the phenomenon of contextuality in a number of other situations
(of the KS--type).
 Moreover, this
alternative is supported by the empirical fact that past, present and
future are different.
The validity of CFD at that level would mean
that nature allows history to be repeated.

\section{COMPARISON WITH STAPP'S RECENT NONLOCALITY ARGUMENTATION}
\label{stapajp031}
\subsection{Equivalence of Stapp's Property
``Free Choices" with CFD and HVs}
Comparing Stapp's earlier proofs
(e.g.\ Refs.\ 9, 44)
with his more recent ones (Refs.\ 10, 11)
Stapp replaces
his former assumptions HV and CFD by a property called ``Free
Choice". He claims that CFD and HV entail ``Free Choice", but that the
converse is not true. We argue, however, that in the
context of his nonlocality argumentation, ``Free Choice" is
equivalent with CFD and HV.

As mentioned in Sec.\ 5.1,
Stapp tries to get a justification for the
validity of CFD in individual cases
by referring to Bohr and he states:
``This `Free Choice' assumption is
important because it allows the causal part of cause--and--effect
relationships to be identified: {\em the choices made by experimenters to
be considered to be causes}. This identification underlies all Bell--type
arguments about causal relationships". Therefore, in his argumentation
``CFD" is now replaced by ``Choice". It follows, however, from our
discussion in Sec.\ 4.1 that by reasoning on individual
quantum events and processes in terms of ``choice", the existence of the
notions of HV and of HVT is still implicitly assumed.
Because only one choice can be made at a time, each choice (made at
one particular instant) must be represented by its own a HV state
in the appropriate HVT.
In nonlocality
argumentations, the individual case concerns an entangled situation, and
HV states $\lambda_L,\lambda_R$
represent the physical situation in two spacelike separated
regions $L$ and $R$. Stapp's
idea to introduce choice (e.g.\ in region $R$) implicitly assumes, then,
that the HV state $\lambda_L$
remains identically the same at different times.
But this is again a counterfactual
statement about an individual quantum process which,
as follows from our discussion in Sec.\ 5,
is only justified  under the condition that RH is valid.
Hence, the notions of ``Choice"  and
``CFD", or ``RH", are identical when translated in
the necessary HV formalism.

\subsection{Stapp's Recent Am.\ J.\ Phys.\ Argumentation}
Stapp's argumentations in Refs.\ 10, 11
are based
on the Hardy--state
\begin{equation}
\ket{\Psi}=N(\ket{c_1}\ket{c_2}-A^2\ket{u_1}\ket{u_2})
\label{hardystateorig}
\end{equation}
(Eq.\ (12) in Ref.\ 45)
which is a correlated state for two systems
$i=1,2$, with $\{\ket{c_i},\ket{d_i}\}$, $\{\ket{u_i},\ket{v_i}\}$
different sets of basis vectors, corresponding with incompatible
observables for both systems. $N$ is a normalization constant, and the
choice of the constant $A^2$ is such that $A^2=\scpr{u_1u_2}{c_1c_2}$.
Stapp denotes these incompatible observables by $L1,L2$ in the
left region $L$, with possible outcomes $L1\pm,L2\pm$, and  by $R1,R2$ in
the right region $R$, with possible outcomes $R1\pm,R2\pm$. The
correspondence between both notations is given in Stapp's Appendix A:
\begin{align}
L1(+,-) &\rightarrow (c_1,d_1) \\
L2(+,-) &\rightarrow (v_1,u_1) \\
R1(+,-) &\rightarrow (d_2,c_2) \\
R2(+,-) &\rightarrow (u_2,v_2).
\end{align}
In this notation (\ref{hardystateorig}) becomes:
\begin{equation}
\ket{\Psi}=N(\ket{L1+,R1-}-\ket{L2-,R2+}
\scpr{L2-,R2+}{L1+,R1-})
\label{hardystatestapp}
\end{equation}
and obeys the relations:
\begin{align}
\scpr{L2-,R2+}{\Psi}=0 \label{orthon1}\\
\scpr{L1-,R2-}{\Psi}=0 \label{orthon2}\\
\scpr{L2+,R1+}{\Psi}=0 \label{orthon3}\\
\scpr{L1-,R1+}{\Psi}\neq 0. \label{nonorthon}
\end{align}
With (\ref{orthon1}), (\ref{orthon2}), (\ref{orthon3})
there correspond
the following predictions with certainty:
\begin{align}
L1- &\Rightarrow R2+ \label{pred1}\\
R2+ &\Rightarrow L2+ \label{pred2}\\
L2+ &\Rightarrow R1- \label{pred3},
\end{align}
while (\ref{nonorthon}) corresponds with
\begin{equation}
L1- \nRightarrow R1- \label{pred4}.
\end{equation}
As Mermin\cite{12} remarks,
the predictions corresponding with (\ref{pred1}),\ldots
,(\ref{pred4}) should be considered as facts which are
independent from the reference frame in which the experiments
are described. This, then, should justify Stapp to set up the chain
\begin{equation}
L1- \Rightarrow R2+ \Rightarrow L2+ \Rightarrow R1-,
\end{equation}
which, however, contradicts (\ref{pred4}), allowing Stapp to
complete his nonlocality argumentation. As mentioned in the Introduction,
Mermin's criticism was that at some stage of the complete proof
in Ref.\ 11, CFD was used in an unjustified way.

Now, for the state (\ref{hardystatestapp}),
Stapp considers in his new argumentation
a Lorentz frame LF in which the observed outcomes in $L$
occur before those in $R$. He
arrives then at his nonlocality conclusion by proving
two Properties 1 and 2.
For the proof he uses the predictions with certainty
(\ref{pred2}) and (\ref{pred3}).
With the notation $L2$ meaning that experiment $L2$ is performed in $L$,
Property 1 reads:\\
``Property 1. Quantum theory predicts that if an experiment of the
Hardy--type is performed then, $L2$ implies $SR$, where,
\begin{itemize}
\item[$SR$] = If $R2$ is performed and gives outcome $+$, then if,
instead, $R1$ had been performed the outcome would have been $-$."
\end{itemize}
The proof of this property is based on the assumption that ``\ldots the
choice made in $R$ does not affect the outcome that has already occurred
in $L$\ldots". Here Stapp is definitely reasoning with {\em single}
quantum events which in general are not described by QM, i.e.\ one has
gone outside QM's domain of applicability. This means that the predictions
with certainty (\ref{pred2}) and (\ref{pred3}) are valid only for two {\em
different} actual ensembles, $\me_1,\me_2$, each of which is described by
the same Hardy--state (\ref{hardystatestapp}). However, using the same
predictions with certainty for {\em one single} ensemble, $\me_1$ or
$\me_2$, has no sense within QM.

Indeed, from our discussion in
Sec.\ 4, it follows that
a rational quantitative account
of individual quantum phenomena
presupposes the availability of a HVT. The only exceptions
are those cases where QM makes predictions
{\em with certainty}, i.e.\ for each of the predictions
(\ref{pred1}), (\ref{pred2}), (\ref{pred3}) {\em separately}
in ensembles $\me_1,\me_2,\me_3$.
However, the point is that Stapp {\em combines} such
predictions with certainty
for {\em one single} ensemble. Our argument is that
a description of such a situation presupposes a HVT in
terms of HV states.
Furthermore,
if one wants to set up a nonlocality argumentation,
our criteria CV and CA require that
within that theory it should be possible to keep the cause for the
occurrence of the earlier outcome in region $L$ constant. In HV
terms this means that the HV state $\lambda_L$ in $L$ (which gives
rise to the occurrence of the definite outcome $r_L$ in $L$
(e.g.\ $L1\pm$ or $L2\pm$)) should be
{\em reproducible}. Because $R1$ and
$R2$ are incompatible, the {\em actual} outcomes
in $R$ and $L$ can only be
obtained in {\em different} experiments, occurring at {\em
different} times $t_{R1}$ and $t_{R2}$.
For these times, one should consider
in the HVT different HV states
$\lambda_L(t_{R1})$ and $\lambda_L(t_{R2})$. Now, Stapp's
assumption that a free choice in $R$ which does not influence
the past outcome $r_L$ in $L$, amounts to the condition
that the cause for the occurrence of the actual outcome,
namely the HV $\lambda_L$, can be kept constant and gives rise to the
result $r_L$, i.e.:
\begin{equation}
\lambda_L(t_{R1})=\lambda \longrightarrow r_L, \quad
\lambda_L(t_{R2})=\lambda \longrightarrow r_L.
\end{equation}
Only this assumption allows Stapp to use his results
(\ref{pred2}), (\ref{pred3}) in tandem
to get his Property 1.

It is seen that in our quantitative approach, the justification of
Stapp's counterfactual statements in his property $SR$ (which is in
terms of observations in $R$),
depends crucially on the reproducibility of the HV state $\lambda_L$
as the cause in
the region $L$ where the actual outcome $r_L\equiv L2+$
of experiment $L2$ is obtained.
Hence, it is {\em not} the assumption that it is the choice in
$R$ that must be considered as the cause for the constancy of $L2+$,
but rather the identity of the HV states
$\lambda_L(t_{R1}),\lambda_L(t_{R2})$.
Only with the assumption of this identity of HV states,
i.e.\ their reproducibility, may Stapp's argumentation be carried on
to the inconsistency with his Property 2, from which his
conclusion to the existence of
``\ldots {\em some sort of} faster--than--light influence" follows.

As a consequence, if the assumption RH leads to the nonlocality conclusion
(which we reject on empirical and theoretical grounds, see Sec.\ 3,
the origin of contradictory conclusions should be the
invalidity of RH. Hence, without the assumption RH, Stapp's premise $SR$
is physically irrelevant. Our present criticism is general, and applies
equally well to all former nonlocality argumentations.

\section{LOCAL HV SCHEME UNDERLYING QT}
QM answers successfully all relevant questions, and covers present day
technological capabilities. Hence, there is as yet no empirical need for a
more detailed HVT, and no one has the slightest idea how to start such a
program. However, some theoretical arguments for yet to consider
the idea to base QM on a HVT are 1) the dissatisfaction with the present
statistical QM, 2) the attempts to remove the alleged nonlocality property
of QM, and 3) the ontological
existence of EPR--type ``elements of reality" which
have no represention in QM.

A convincing proof of this
existence is given by Renninger,\cite{39} by means of
an interferometer thought experiment.
Using such a setup, Renninger showed unambiguously that
{\em with each single quantum system moving through the interferometer}
there correspond realities moving along both paths. In terms of the
notions particle and wave, one path should contain the particle, and the
reality in the other path is usually called an ``empty wave" (EW). By
incerting phase shifters (in fact, ``half wavelength" plates) he showed
explicitly that these realities, including the empty wave, are {\em
causally influencable} in each single case.
 In contrast with what Stapp calls ``orthodox quantum
philosophy", Renninger stated that ``Die Anschaulichkeit soll nicht
vorzeitig aufgegeben werden", i.e.\
``The pictorial representation should
not be given up prematurely".
Of course, these
cases are covered well
by QM because the interventions in both paths
correspond with predictions with certainty and, hence, apply to
each individual case. Recently, Hardy\cite{46}
has tried to prove the converse, namely that also the EW is able to change
the state of a quantum system. It may be shown\cite{47}
that Hardy's argumentation is based on
assumptions which lead to contradictory
QM results.
However,
notwithstanding the present lack of evidence for
EWs influencing
themselves directly and observably other systems,
Renninger has shown unambiguously their ontologic existence
(see also Sec.\ 4.3).

Because all available empirical data
are described very well by QM, they
cannot be used as a clear guide
for introducing a new quantitative scheme, as a substitute for QM.
Therefore,
we are only able to give a very general account of
our resolution of the so--called ``incompatible
quantum results". In agreement with our starting points above, we shall
adhere to the view that for each single
observation there exists a causal reason
on a deeper level of description. We shall assume that, instead of
chance, it is our {\em ignorance} of the precise individual state,
following a preparation or a measurement, which is responsible for the
uncertainty about future outcomes. As the variables in QM allow only
the deterministic prediction of the statistics, we will need supplementary
(``hidden") variables and a new
(``hidden variable") theory allowing
the prediction of individual outcomes for specific initial conditions.
As remarked above,
it is nature itself that determines which
initial conditions are realizable, and whether they are reproducible or not
in subsequent runs of an experiment.
Such a HVT for individual quantum processes should conform
the general requirements of a theory, namely its principles should agree
with empirical data (such as locality), and the initial conditions should
allow to predict in a causal and deterministic
way the unique outcome of future events, but only one at a time.

Consider, then, two identically prepared ensembles $\me_1,\me_2$, consisting
each of $N$ individual preparations, and described by a quantummechanical
state $\ket{\Psi}$.
Labeling each element in an ensemble by $i$, we may represent $\me_1,\me_2$
also in terms of sets of HV states by
$\left\{ \lambda(x^{\mu}_{i;1}),i=1,\ldots,N\right\}$, and
$\left\{ \lambda(x^{\mu}_{i;2}),i=1,\ldots,N\right\}$,
where dependence on space--time coordinates $x^{\mu}$
is explicitly noted.
In
this way we may write the following equivalences:
\begin{equation}
\left\{ \lambda(x^{\mu}_{i;1}),i=1,\ldots,N\right\} \sim
\ket{\Psi}\;,
\end{equation}
and
\begin{equation}
\left\{ \lambda(x^{\mu}_{i;2}),i=1,\ldots,N\right\} \sim
\ket{\Psi}\;.
\end{equation}
We now remark that the identity with respect to the QM state
vectors does not necessarily mean identity with respect to the HV
states and, hence, does not imply
\begin{equation}
\lambda(x^{\mu}_{i;2}) = \lambda(x^{\mu}_{i;1}), \,i =
1,\ldots,N\;,
\end{equation}
i.e.\ the sets of individual states
\begin{equation}
\left\{ \lambda (x^{\mu}_{i;1}), i=1,\ldots,N\right\},
\quad
\left\{
\lambda (x^{\mu}_{i;2}),i=1,\ldots,N\right\}
\end{equation}
are not necessarily identical, even after any possible
reorganization of the elements of the sets.
The possibility that such nonidentical ensembles on the HV level,
may yet give rise to identical QM predictions for the statistics, has been
proven by Kupczynski.\cite{48}

In the domain of application of CM and of QM, nature allows the
preparation of physical situations such that their mathematical
representation is by identical functions, i.e.\ the corresponding physical
states are reproducible. However, it is not evident that nature allows the
same to be true for {\em individual} HV states, i.e.\ this remains a
supplementary assumption which may be invalid at a HV level of
description. In the above sections we have argued in favour of the
nonreproducibility of physical states at the HV level of description. As
is the case with Einstein locality, this nonreproducibility is in full
agreement with the empirical fact that past, present and future are
observably distinct. Hence, ``nonreproducibility of individual HV states"
may be a possible solution for saving locality, and for removing all
quantum paradoxes.

\section{CONCLUSIONS}
We have started from the observation that all known empirical data
verify the principle of Einstein locality, as defined in
Sec.\ 2.
We have argued that, rather than questioning that principle,
it should be incorporated firmly in our physical theories.
This has been
shown the case for QM and its more recent elaborations.

Assuming the
validity of this principle also in a tentative HVT
for {\em individual} quantum processes, we have come to the conclusion
that, in agreement with the empirical evidence that past, present and future
are basically different,
the validity of the notion of CFD breaks down at the HV level.
In physical terms this means
that HV states $\lambda(x^{\mu})$ may have the property of being
``nonreproducible''. In other words,
one may envisage the situation that the HVT's initial conditions can no
longer be reproduced, so that it no longer has sense to ask
for a single event the
counterfactual question ``what would have happened if, instead of
the actual choice, another choice had been made''.

We have argued that
(formal) determinism and causality may still be valid
in a HVT, just like in CM and in QM. Furthermore, specifying to
{\em correlated} observations, the notions {\em cause} and
{\em effect} apply between the preparation and the correlated
measurements, but {\em not} between the correlated measurements
themselves, the time sequence depending on the frame of
reference. The existence
of a correlation may be explained simply on the ground that
there is a common past.

We conclude further that our NRH is
compatible with the empirical existence of an arrow of time, and
agrees with statements such as one of Y. Aharonov, P.G. Bergmann,
J.L. Lebowitz:\cite{49} ``One of the perenially challenging
problems of theoretical physics is that of the `arrow of time.' Everyday
experience teaches us that the future is qualitatively different from the
past, that our practical powers of prediction differ vastly from those of
memory, and that complex physical systems tend to develop in the course of
time in patterns distinct from those of their antecedents."

Finally, in our
approach we have got a reconciliation between Einstein's and Bohr's views.
EPR's suggestion that QM may be completed by a more detailed local HVT is
shown to be possible. In such a theory, EPR's ``elements of
physical reality" should correspond with time--dependent, but
nonreproducible, initial conditions in the HVT. In this way EPR's
conclusion of the simultaneous existence of quantummechanical incompatible
observables can be avoided. This solution roughly corresponds with EPR's
last paragraph, in which they offer a similar way out of their
contradictory conclusion of the simultaneous existence of
quantummechanically incompatible elements of reality. Because the
evolution of physical processes may proceed
according to deterministic laws,
Einstein's dictum that ``God does not play dice" is met.

If physics is defined as being concerned, as Bohr would have it,
with observable, {\em reproducible}, phenomena then a HVT, which
would describe the dynamics of $\lambda(x^{\mu})$, would be, at
least at present, of no practical use because of the basic
failure to know exactly the initial conditions. In that case it
will be difficult -- if not impossible for human beings --
to surpass QM, which then may be considered a
complete 
``FAPP"\footnote{FAPP is short for ``For All Practical Purposes'',
and has been used by Bell\cite{50} and much earlier by 
Carruthers and Nieto.\cite{51}}
theory because it answers all possible questions
which may ever be posed by human observers. In this case it is not God
himself who plays dice, but rather the human
observer because of his intrinsic limitations.

Hence, from a fundamental point of view EPR are right. However,
those prefering a more pragmatic point of view, may join Bohr in that
it may turn out impossible, because of the observer's limitations,
to describe causally and quantitatively
individual quantum processes. Yet, surprises may never be excluded,
and in a near future one may get more control over an objectively
existing reality, such as brought about by the
Renninger thought experiment.

\noindent
\section*{ACKNOWLEDGEMENT}
The author acknowledges constructive comments by one referee,
which lead to the present version of the paper.

\end{document}